\documentclass[10pt,showpacs,twocolumn,aps,pre,groupeaddress]{revtex4}

\usepackage[pdftex]{graphicx}
\usepackage{mathrsfs}
\usepackage{amsmath}
\usepackage{amssymb}
\usepackage{amsfonts}
\usepackage{ae}
\usepackage{units}
\usepackage{appendix}

\begin{document}
%
\title{Dynamics of a suspended nanowire driven by an ac Josephson current in an inhomogeneous magnetic field}

\author{Milton E. Pe\~na-Aza}
\affiliation{Department of Physics, University of Gothenburg, SE-412 96 G\"oteborg, Sweden}
\date{\today}
%
\begin{abstract}

We consider a voltage-biased nanoelectromechanical Josephson junction, where a suspended nanowire forms a superconducting weak-link, in an inhomogeneous magnetic field. We show that a nonlinear coupling between the Josephson current and the magnetic field generates a Laplace force that induces a whirling motion of the nanowire. By performing an analytical and a numerical analysis, we demonstrate that at resonance, the amplitude-phase dynamics of the whirling movement present different regimes depending on the degree of inhomogeneity of the magnetic field: time independent, periodic and chaotic. Transitions between these regimes are also discussed. 

\end{abstract}

\maketitle


\section{INTRODUCTION}
\label{section1}

Over the past fifteen years, nanoelectromechanical systems (NEMS) have been widely used for the exploration of the quantum world and for the development of new technological applications \cite{Craighead2000, Roukes2001b, Roukes2001, clelandbook, cho2, ekinci, poot}. Because of their size, high frequency operation, small mass and high performance, NEMS are currently considered promising candidates for achieving the quantum limit of mechanical motion. It is expected that the quest for the quantum regime in such devices will elucidate questions of fundamental nature in physics, for instance, the quantum-mechanical description of  macroscopic objects \cite{legget, blencowe, roukes, PhysRevLett.99.040404, cho, aspelmeyer}. Similarly, physical systems involving NEMS resonators are excellent tools for theoretical and experimental studies of nonlinear dynamical systems \cite{raro, rhoads}. Indeed, examples of complex dynamical phenomena in NEMS are numerous and include chaotic behavior \cite{Karabalin2009,arabe1, arabe2, Kenig2011}, bifurcation-topology amplification \cite{PhysRevLett.106.094102}, nonlinear switching dynamics \cite{PhysRevB.81.241405} and nonlinear frequency pulling \cite{PhysRevLett.93.224101} to name but a few.

By making use of the potential offered by NEMS resonators, G. Sonne \textit{et al.}~\cite{PhysRevB.78.144501} studied the nonlinear dynamics of a suspended carbon nanotube coupled to two voltage biased superconducting electrodes. In presenting their work, Sonne and collaborators assumed that the nanoelectromechanical junction was subjected to a \textit{homogeneous} magnetic field perpedicular to the axes of the nanowire. For such a system, the authors demonstrated the possibility to pump energy from the electronic subsystem into the mechanical vibrations; they also demonstrated that the system had more than one regime of finite-amplitude stationary nonlinear oscillations. In particular, a region of bistability was found and the authors showed that it should be detected in the corresponding dc Josephson current (see discussion in Ref. 23).

In this article, we consider the same voltage-biased nanoelectromechanical system studied by Sonne and co-workers \cite{PhysRevB.78.144501}, but now  we extend the analysis to the case in which the NEMS resonator is subjected to a \textit{nonuniform} magnetic field. As will be discussed below, inhomogeneity of the field causes the conducting nanoresonator to execute a whirling movement resembling a jump-rope like motion. The purpose of this paper is to analyze the time evolution of the amplitude and relative phase of the nanotube whirling motion. We will demonstrate that the coupled amplitude-phase dynamics exhibit different stationary regimes depending on the degree of the magnetic field inhomogeneity: time independent, periodic and chaotic.

The scope of this article is parallel to research conducted by Conley \textit{et al.}~\cite{conley} and Chen \textit{et al.}~\cite{chen}, who previously investigated nonlinear and nonplanar dynamics of suspended nanowires excited by an electrostatic force. Here, however, we consider another type of driving mechanism: the Laplace force. This excitation mechanism to accomplish  whirling motion of suspended nanoresonators has not been considered before.

The article is organized in the following manner. In Sec.~\ref{section2}  we introduce the model Hamiltonian for the considered system and examine the coupling between the mechanical and electronic subsystems. At the end of the section, we derive the equations of motion governing the dynamics of the nanoresonator driven by an ac Josephson current. In Sec.~\ref{section3} we obtain the differential equations for the time evolution of the amplitudes of the nanotube vibration and  discuss their dynamics through the analysis of the numerically computed results. Finally, in Sec.~\ref{final} we provide the summary and concluding remarks of this work.


\section{MODEL HAMILTONIAN SYSTEM}
\label{section2}

\begin{figure}
\includegraphics[width=8.5cm]{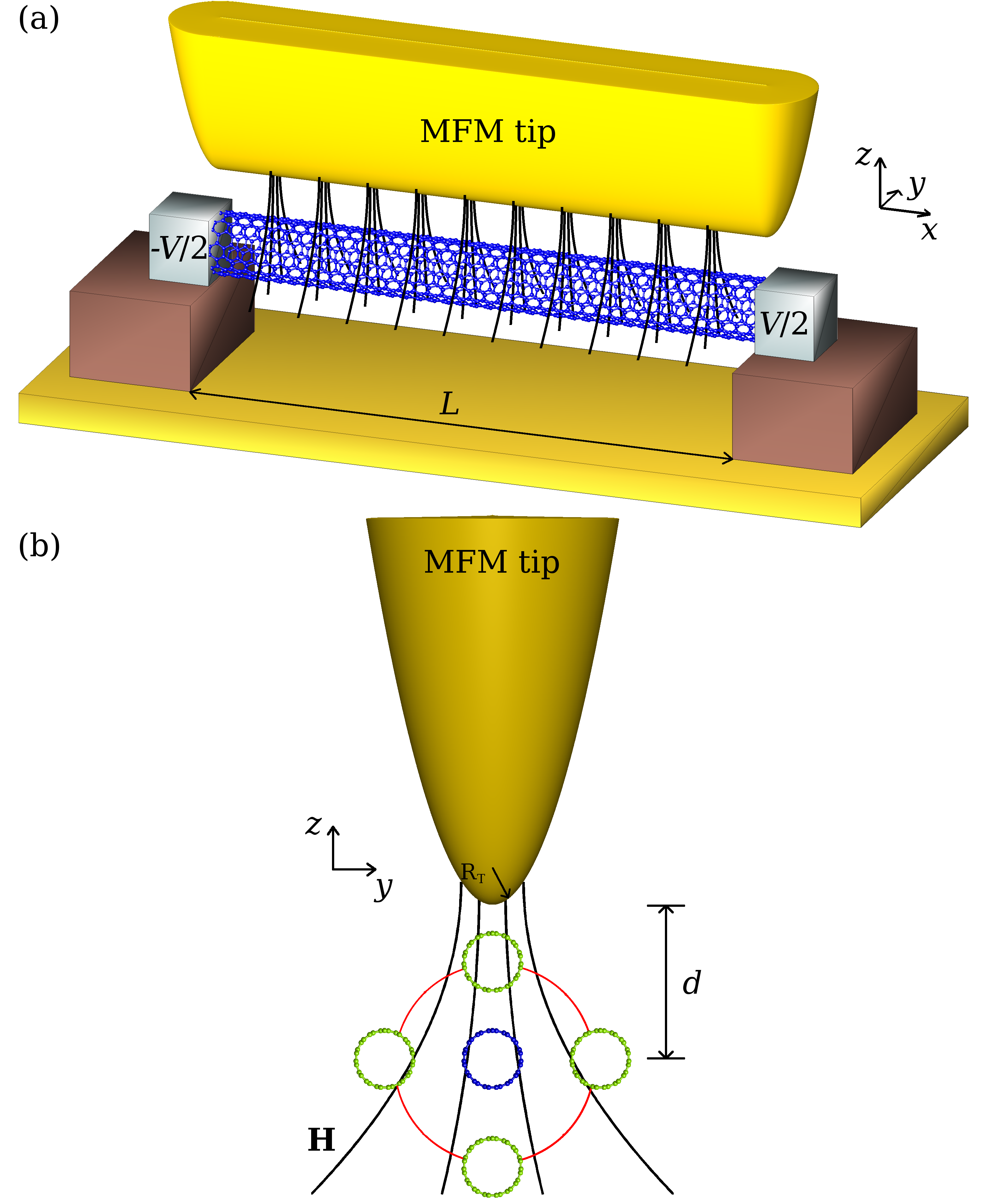}
\caption{(Color online) (a) Schematic diagram of the voltage-biased S-N-S nanoelectromechanical Josephson junction
 considered in the article. A doubly clamped metallic carbon nanotube suspended over a trench of length $L$,
  forms a weak link between two voltage-biased superconducting electrodes. The junction is influenced by an
  inhomogeneous magnetic field $\mathbf{H}$, generated by a wedge-shaped MFM cantilever tip  at a distance $d$ from the nanotube at rest (blue circle). (b) Nanotube displacement (green circles) in the $z$-$y$ plane induced by a Laplace force. $R_{T}$ is the curvature radius of the magnetic tip.}
\label{carbon}
\end{figure}

The diagram in Fig.~\ref{carbon}  is a schematic illustration of a superconducting hybrid nanostructure, a superconducting-normal-superconducting (S-N-S) nanoelectromechanical Josephson junction driven by a dc voltage bias $V$. The junction consists of a metallic carbon nanotube of length $L$ suspended between two voltage-biased superconducting leads. In such a geometry, the nanotube is simultaneously serving as a mechanical resonator and as a weak link between the superconducting electrodes. In the layout of the system, the NEMS junction is under the influence of an external inhomogeneous magnetic field $\mathbf{H}$, generated by a magnetic force microscope (MFM) cantilever tip parallel to the axis of the nanotube at a distance $d$.

As demonstrated by Sonne \textit{et al.}~\cite{PhysRevB.78.144501}, the electronic subsystem can be effectively modeled in terms of the Bogoliubov-de Gennes Hamiltonian
\begin{subequations}\label{eq1}
\begin{gather}
\hat{\mathcal{H}} = \int{\boldsymbol{\hat{\Psi}}^\dagger(x)\bigl(\hat{\mathcal{H}}_0 + \hat{\mathcal{H}}_\Delta\bigr)\boldsymbol{\hat{\Psi}}(x)dx}\,, \label{eq1a}\\
\hat{\mathcal{H}}_0 = -\frac{\hbar^2}{2m}\hat{\sigma}_z\biggl(\frac{\partial}{\partial x} - \hat{\sigma}_z\frac{ieA_{x}}{\hbar}\biggr)^2 + \hat{\sigma}_{z}U(x)\,,\label{eq1b}\\
\hat{\mathcal{H}}_{\Delta} = \Delta(x)\bigl[\hat\sigma_{x}\cos \phi(t) + \mathrm{sgn}(x)\hat{\sigma}_{y}\sin \phi(t)\bigr]\,.\label{eq1c}
\end{gather}
\end{subequations}
Here, $\boldsymbol{\hat{\Psi}}(x)$ and its complex conjugate  $\boldsymbol{\hat{\Psi}}^\dagger(x)$ are two-component Nambu spinors that annihilate and create electrons/holes at position $x$, respectively. In the Bogoliubov-de Gennes equation, $\hat{\mathcal{H}}_0$ is the Hamiltonian for electrons/holes in the metallic carbon nanotube, where $U(x)$ is the potential barrier between the tube and the bulk superconducting leads, $A_{x}$ is the magnetic vector potential in the $x$-direction and $\hat{\sigma}_{i}$ are the Pauli matrices. The superconducting electrodes are described by $\hat{\mathcal{H}}_{\Delta}$, where the gap function is given by $\Delta(x)= \Delta_{0}\Theta(2|x|-L)$, with $\Delta_{0}$ the order parameter and $\Theta(x)$ the Heaviside step function. The phase difference between the superconducting leads is denoted by $\phi$ and, in accordance to the second Josephson relation, its time evolution is $\phi(t)= \omega_Jt$, with $\omega_{J}=2eV/\hbar$ the Josephson frequency.

We now consider a  magnetic field of the form $\mathbf{H}=(0, H_{y},H_{z})$, which can be generated, for instance, by an MFM cantilever tip in the form of a wedge, \emph{cf.} Fig.~\ref{carbon}.  A first order Taylor expansion of the magnetic field yields $\mathbf{H}=\left(0,y\partial_{y}H_{y}(x,0,0), H_{z}(x,0,0) + z\partial_{z}H_{z}(x,0,0)\right)$, where  $H_{z}(x,0,0)$ and $\partial_{i}H_{i}(x,0,0)$ represent the magnitudes of, respectively, the magnetic field $z$-component and the magnetic field gradients, both evaluated at the axis of the nanotube.  A straightforward calculation from the Maxwell equation $\boldsymbol{\nabla} \cdot \mathbf{H}=0$, indicates that $\partial_{y}H_{y}(x,0,0)=-\partial_{z}H_{z}(x,0,0)\equiv-H'_{z}$. Setting $H_{z}(x,0,0)\equiv H_{z}$, the magnetic field reads
\begin{equation}
\mathbf{H}=\left(0,-H'_{z}y, H_{z} + H'_{z}z\right)\,.
\label{magnetifield}
\end{equation}
This field is also obtained from the equation $\mathbf{H}=\boldsymbol{\nabla}\times \mathbf{A}$, where the vector potential $\mathbf{A}$ is given by
\begin{equation}
\mathbf{A}=-\bigl(\bigl[H_{z} + H'_{z}z\bigr]y,0,0\bigr)\,.
\label{eq2}
\end{equation}
\begin{figure*}
\includegraphics[width=14cm]{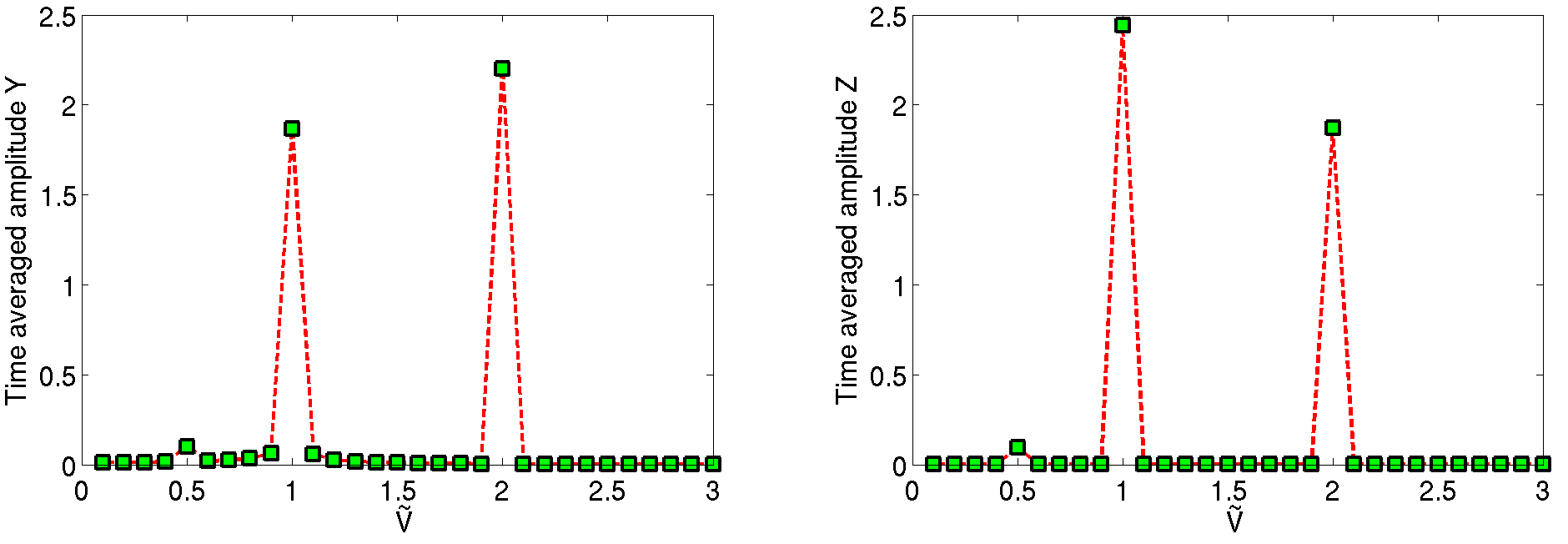}
\caption{(Color online) Numerical simulations of  Eqs.~{\eqref{eq4a}} and {\eqref{eq4b}} for the time averaged coordinates $Y$ and $Z$  as function of the driving voltage $\tilde{V}$. The plots are calculated for $\tilde{\gamma}=0.001$, $\varepsilon=0.012$, $\varkappa=1$ ($L=$~\unit[1]{$\mu$m}, $h_z=$~\unit[40]{mT}, $R_{T}=$~\unit[66]{nm}).}
\label{amplitudes1}
\end{figure*}
By substituting the magnetic vector potential, Eq.~{\eqref{eq2}}, into the Hamiltonian given by Eq.~{\eqref{eq1b}}, we can readily observe that $\hat{\mathcal{H}}_0$ is altered by deflecting the nanotube in the $z$-$y$ plane. In this case, non-planar whirling displacement of the nanotube is due to the Laplace force generated by coupling between the magnetic field and the Josephson current. As a consequence of the motion in the magnetic field, an electromotive force is induced along the nanomechanical weak-link and its magnitude depends on the rate of change of the nanowire profile in the $z$-$y$ plane, \emph{i.e.}, the rate of change of the magnetic flux through the circuit. In this description, now, the superconducting phase difference is not only a function of the bias voltage, but also of the nanowire deflection in the $z$-$y$ plane. We decompose the nanotube motion in this plane into two independent deflections $y(x,t)=u_{0}(x)a(t)$ and $z(x,t)=u_{0}(x)b(t)$, where $u_{0}(x)$ is the normalized and dimensionless profile of the fundamental mode in both directions. Then, following Shekhter \emph{et al.}~\cite{PhysRevLett.97.156801}, a unitary transformation which moves the vector potential from the kinetic part of $\hat{\mathcal{H}}_0$ to the phase difference between the superconducting electrodes is applied. As a result, the expression for the superconducting phase difference has the form
\begin{equation}
\varphi(t)=\frac{\phi(t)}{2} + \frac{4eL}{\hbar}\bigl[h_{z}+ h'_{z}b(t)\bigr]a(t)\,.
\label{eq3}
\end{equation}
The parameters $h_{z}=\alpha H_{z}$ and $h'_{z}=\beta H'_{z}$ are the renormalized magnetic field and magnetic field gradient in the $z$-direction calculated at the axis of the tube, respectively, with $\alpha,\beta\thicksim 1$ correctional factors originating from geometrical considerations.

The nanotube mechanics is thus described through the projection amplitudes $a(t)$ and $b(t)$.  For the conjugate variables $\{a(t),p_{a}(t)\}$ and $\{b(t),p_{b}(t)\}$ ($p_{j}(t)$ denoting the generalized momenta) one can formulate the following Hamiltonian function   
\begin{multline}
H(p_{a},p_{b},a,b,t)=\frac{1}{2m}\left(p^{2}_{a} + p^{2}_{b}\right) +\frac{m\omega^{2}}{2}\left(a^2 + b^{2}\right) - \\
- 2D\Delta_{0}\cos(\varphi(a,b,t))\,,
\label{hamilton}
\end{multline}
where $m$ and $\omega$ are the mass and the mechanical eigenfrequency of the nanoresonator, respectively. The last term in Eq.~{\eqref{hamilton}} corresponds to the Josephson energy $E_{J}(\varphi(a,b,t))$, with $D$ the transmission coefficient of the junction. The equations of motion for $a(t)$ and $b(t)$ are then obtained from the Hamilton equations. Written in terms of the dimensionless deflection coordinates $Y(t)=\bigl[4eLh_{z}\bigr/\hbar]a(t)$ and $Z(t)=\bigl[4eLh_{z}\bigr/\hbar]b(t)$, the resulting set of differential equations for the nanotube amplitudes is:
\begin{subequations}
\label{eq4}
\begin{multline}
\ddot{Y}(\tau) + \tilde{\gamma}\dot{Y}(\tau) + Y(\tau)  =\\
 -\varepsilon\left[1+\varkappa Z(\tau)\right]\sin\left(\tilde{V}t+\left[1+\varkappa Z(\tau)\right]Y(\tau)\right)\,\label{eq4a},
\end{multline}
\begin{multline}
\ddot{Z}(\tau) + \tilde{\gamma}\dot{Z}(\tau) + Z(\tau)  = \\
 -\varepsilon\varkappa Y(\tau)\sin\left(\tilde{V}t+\left[1+\varkappa Z(\tau)\right]Y(\tau)\right)\,\label{eq4b}.
\end{multline}
\end{subequations}
Here, we have added a dimensionless phenomenological damping coefficient $\tilde{\gamma}= [\gamma/m\omega$]. In these equations $\varepsilon=[8eL^2h_{z}^2j_c/m\hbar\omega^2]$ with $j_{c}=[D\Delta_0e/2\hbar]$, the critical current through the junction. We also set the timescale to $\tau=\omega t$ and, consequently, $\tilde{V}=[eV/\hbar\omega]$. The parameter 
\begin{equation}
\varkappa=\frac{\hbar}{4eLh_{z}R_{T}}\,, 
\label{eq5}
\end{equation}
where $R_{T} = \bigl[h_{z}/h'_{z}\bigr]$ denotes the curvature radius of the magnetic cantilever tip, characterizes the degree of inhomogeneity of the magnetic field and will be referred to as the \textit{control parameter}. It turns out that $\varkappa$ determines the dynamical behavior of the system. One can then realize the significance of $\varkappa$ by considering fixed system parameters ($L,h_{z}$) and by letting the $R_{T}$ vary. In the limit $R_{T} \to \infty$, the control parameter vanishes and the equations of motion given by Eqs.~{\eqref{eq4}} reduce to the case discused by Sonne \textit{et al.} \cite{PhysRevB.78.144501} where the magnetic field is uniform and, therefore, the amplitude of the driving force acting in the $z$-direction becomes zero. 

\section{NUMERICAL RESULTS AND DISCUSSION}
\label{section3}

For a qualitative and quantitative discussion of the dynamic behavior of the nanowire amplitudes, we consider the following system parameters: a carbon nanotube of radius $r=$~\unit[1]{nm} and length $L=$~\unit[1]{$\mu$m}~\cite{saito}, superconducting order parameter $\Delta_{0}\sim$~\unit[10]{meV}, Josephson critical current $j_{c}\sim$~\unit[100]{nA}~\cite{kasumov}, and quality factor $Q\sim 10^{3}$~\cite{witkamp}. We also assume that $\tilde{\gamma}=1/Q$, $h_{z}\sim$~\unit[40]{mT}, $\varepsilon=0.012$, and $R_{T}$ is varied from \unit[53]{nm} to \unit[544]{nm}.

\begin{figure*}
\includegraphics[width=13cm]{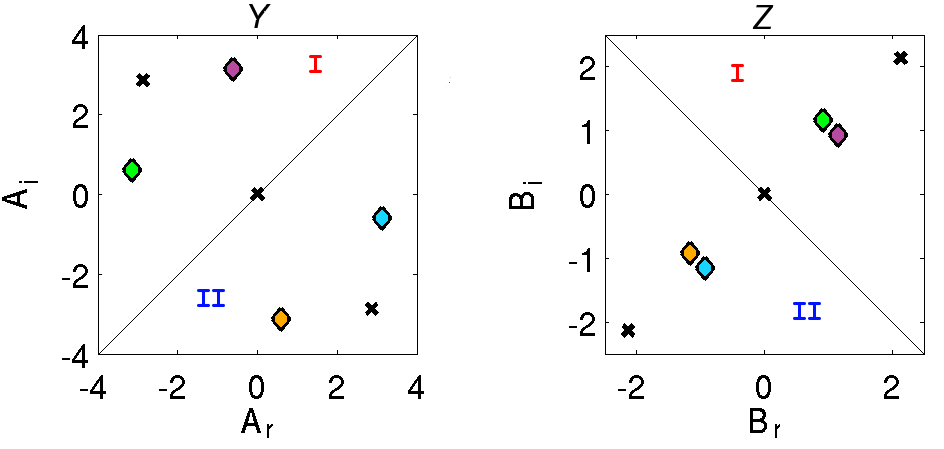}
\caption{(Color online) Numerical simulations for the stability analysis of Eqs.~{\eqref{rwf}}. Stable and unstable stationary points are indicated by colored diamonds ($\Diamond$) and black crosses ($\times$) respectively. Diamonds of identical color indicate the four envelopes of the stable solution $\mathbf{X}_{s}$, crosses in the same region ($\mathtt{I}$ or $\mathtt{II}$) of both phase planes belong to the same unstable solution $\mathbf{X}_{u}$. The zero solution, $\mathbf{X}\equiv \boldsymbol{0}$, is unstable. Here, $\tilde{\gamma}=0.001$, $\varepsilon=0.012$, $\varkappa=0.12$ ($L=$~\unit[1]{$\mu$m}, $h_z=\ $\unit[40]{mT}, $R_{T}=\ $\unit[544]{nm}).}
\label{symmetry2}
\end{figure*}

Numerical simulations of Eqs.~{\eqref{eq4a}} and {\eqref{eq4b}} allow us to study the time average of the nanotube deflection coordinates $Y(t)$ and $Z(t)$ as functions of the driving voltage $\tilde{V}$. In doing so, one can notice that the system response presents a series of resonance peaks at integer values of the driving voltage, \emph{i.e.}, the amplitude of the nanoresonator is not damped provided the resonant condition is fulfilled: the Josephson frequency  matches the mechanical frequency (see Fig.~\ref{amplitudes1}). This resonant phenomenon was first studied by Sonne \textit{et al.}~\cite{PhysRevB.78.144501}, who attributed  a direct resonance at $\tilde{V}=1$  and a parametric resonance at $\tilde{V}=2$. Accordingly, the same conclusion can be drawn from the results presented in Fig.~\ref{amplitudes1}. In the remainder of the article we will be mainly focusing on the parametric regime and take $\tilde{V}=2$. In this case, the dynamic behavior of the amplitudes can be analized by postulating a solution for both deflection coordinates in the form

\begin{subequations}\label{eq8}
\begin{gather}
Y(\tau)=A_{\textit{r}}(\tau)\cos(\tau) + A_{\textit{i}}(\tau)\sin(\tau)\,,\label{eq8a}\\
Z(\tau)=B_{r}(\tau)\cos(\tau) + B_{i}(\tau)\sin(\tau)\,.
\label{eq8b}
\end{gather}
\end{subequations}
On condition that $\tilde{\gamma}, \varepsilon, \varepsilon\varkappa \ll 1 $, the four envelopes in the vector $\mathbf{X}=\left(A_{r}(\tau), A_{i}(\tau), B_{r}(\tau), B_{i}(\tau)\right)$  vary slowly in time, \emph{i.e.}, $ d\mathbf{X}/d\tau \ll 1$ and, an averaging method \cite{libro} can be employed in order to derive the equation of motion for $\mathbf{X}$. By substituting the ansantz provided in Eqs.~\eqref{eq8} into the system of equations in Eqs.~\eqref{eq4} and integrating over the fast oscillations one gets 
\begin{equation}
\begin{aligned}
\frac{dA_{r}}{d\tau} + \frac{\tilde{\gamma}A_{r}}{2}&=\frac{\partial\mathcal{G}}{\partial A_{i}}\,,\\
\frac{dA_{i}}{d\tau} + \frac{\tilde{\gamma}A_{i}}{2}&=-\frac{\partial\mathcal{G}}{\partial A_{r}}\,,
\end{aligned}
\quad\ \   
\begin{aligned}
\frac{dB_{r}}{d\tau} + \frac{\tilde{\gamma}B_{r}}{2}&=\frac{\partial\mathcal{G}}{\partial B_{i}}\,,\\
\frac{dB_{i}}{d\tau} + \frac{\tilde{\gamma}B_{i}}{2}&=-\frac{\partial\mathcal{G}}{\partial B_{r}}\,,
\end{aligned}
\label{rwf}
\end{equation}
where
\begin{multline}
\mathcal{G}(\varkappa)=-\frac{\varepsilon}{2\pi}\int_{-\pi}^\pi\cos\bigl(\left[1+\varkappa B_{r}\cos(\Theta)+\varkappa B_{i}\sin(\Theta)\right]\\
\times\left[A_{r}\cos(\Theta)+A_{i}\sin(\Theta)+2\Theta\right]\bigr)d\,\Theta\,.
\end{multline}
Here, $\mathcal{G}(\varkappa)$ is the generatrix Hamiltonian function. The study proceeds by performing a stability analysis based on Eqs.~\eqref{rwf}. In general, we shall study solutions of a system of coupled ODEs, $d\mathbf{X}/d\tau =\mathbf{f}(\mathbf{X},\varkappa)$, and solutions of a system of algebraic equations $\boldsymbol{0}=\mathbf{f}(\boldsymbol{X},\varkappa)$.

The discussion commences by highlighting the symmetric nature of the stationary solutions for the algebraic system in Eqs.~{\eqref{rwf}}. For the above considered system parameters with $R_{T}=\ $\unit[544]{nm} and $\varkappa=0.12$, numerical simulations for the slowly-varying envelopes show that the system presents several stationary points. Such qualitative behavior is visualized in the phase space diagrams for the nanotube amplitudes, $Y$ and $Z$, in Fig.~\ref{symmetry2}. There, solutions pictured by colored diamonds ($\Diamond$) and black crosses ($\times$) correspond to  stable and unstable stationary points, respectively. Similarly, the phase portraits are divided in two regions, denoted $\mathtt{I}$ and $\mathtt{II}$, by diagonal symmetry axes. The symmetry axes in the $Y$ and $Z$ coordinates satisfy the relations $A_{i}-A_{r}=0$  and $B_{i}+B_{r}=0$, respectively. From Fig.~\ref{symmetry2} it follows that the presented pattern of solutions has a mirror-image symmetry with respect to the line of symmetry that divides the phase diagrams into regions $\mathtt{I}$ and $\mathtt{II}$, where diamonds of the same color represent the four components of a stable computed solution $\mathbf{X}_{s}$ and crosses in the same region ($\mathtt{I}$ or $\mathtt{II}$) of both phase spaces account for the envelopes of an unstable solution $\mathbf{X}_{u}$. From a symmetric point of view, the system is characterized by two stable and two unstable (including $\mathbf{X} \equiv \boldsymbol{0}$) stationary points in each region. However, the dimensionless angular momentum of the nanowire, $\mathbf{L}_{x}=YdZ/d\tau-ZdY/d\tau$,  in both regions is equal in magnitude but opposite in sign.

\begin{table}[b]
\begin{center}
\caption{Bifurcation pattern in the symmetric regions of solutions $\mathtt{I}$ and $\mathtt{II}$ as a function of the control parameter $\varkappa$. Solution $\mathbf{X}\equiv \boldsymbol{0}$ is included.}
\label{table1}
\begin{tabular}{ c p{2.5cm} p{2.5cm} }\hline\hline\multicolumn{1}{c}{} & \multicolumn{1}{c}{Number of} & \multicolumn{1}{c}{Number of} \\ 
\multicolumn{1}{c}{$\varkappa$-interval} & \multicolumn{1}{c}{stable solutions} & \multicolumn{1}{c}{unstable solutions} \\ 
\hline
\multicolumn{1}{c}{$\varkappa \leq 0.120$} & \multicolumn{1}{c}{2} & \multicolumn{1}{c}{2} \\ 
\multicolumn{1}{c}{$0.120 < \varkappa \leq 0.125$} & \multicolumn{1}{c}{3} & \multicolumn{1}{c}{3} \\
\multicolumn{1}{c}{$0.125 < \varkappa \leq 0.140$} & \multicolumn{1}{c}{1} & \multicolumn{1}{c}{3} \\ 
\hline\hline
\end{tabular}
\end{center}
\end{table}  

\begin{figure*}
\includegraphics[height=13.6cm]{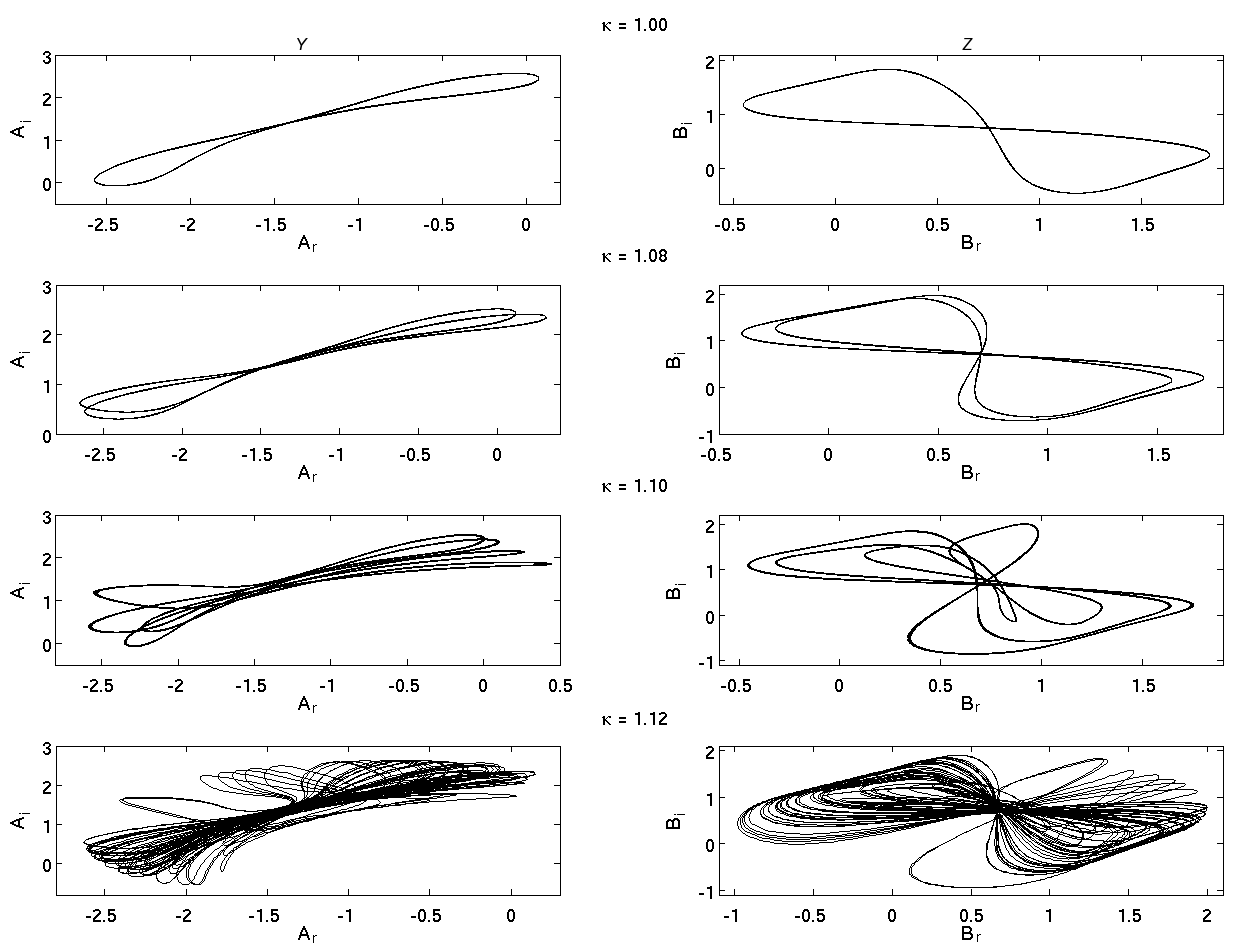}
\caption{Period-doubling bifurcations for different values of the control parameter. Plots are shown for the first symmetry region of both phase spaces. Simulations were obtained for $\tilde{\gamma}=0.001$, $\varepsilon=0.012$ ($L=$~\unit[1]{$\mu$m}, $h_z=$~\unit[40]{mT}). }
\label{bifurcations}
\end{figure*}

Continuing our exploration, the computed results indicate that the bistable regime is only attained for  $\varkappa \leq 0.120$. By letting  the control parameter increase further, bistability is abandoned and the number of stationary solutions of Eqs.~\eqref{rwf} is modified as well as their stability. Indeed, Table~\ref{table1} presents the number of stable and unstable stationary solutions in the interval $ 0.120 \leq \varkappa \leq 0.140$ in both regions for these equations. As can be seen from the table, the system displays a phenomenon called \textit{branching} or \emph{bifurcation}, which is a distinctive fingerprint of nonlinear dynamical systems \cite{seidel}. The exhibited branching pattern formation can be described as follows. For $\varkappa \leq 0.120$,  bistability is accompanied by two unstable stationary points, one of them corresponds to $\mathbf{X}\equiv \boldsymbol{0}$ (see Fig.~\ref{symmetry2}). Once the control parameter is slightly increased over the threshold value $\varkappa = 0.120$, the nonzero unstable stationary point becomes stable, and simultaneously, two new unstable solutions appear and the system has now three stable and three unstable (including the trivial solution) stationary points. Next, by varying the control parameter in the interval $0.120\leq\varkappa \leq 0.125$, the two original stable points and the two unstable ones move in phase space and approach each other. Finally, when the control parameter ranges fom $0.125$ to $0.140$, the bistable and the created unstable solutions coalesce into two unstable solutions. In this stage, there is only one stable and three unstable stationary points.

\begin{figure*}
\includegraphics[width=13.6cm]{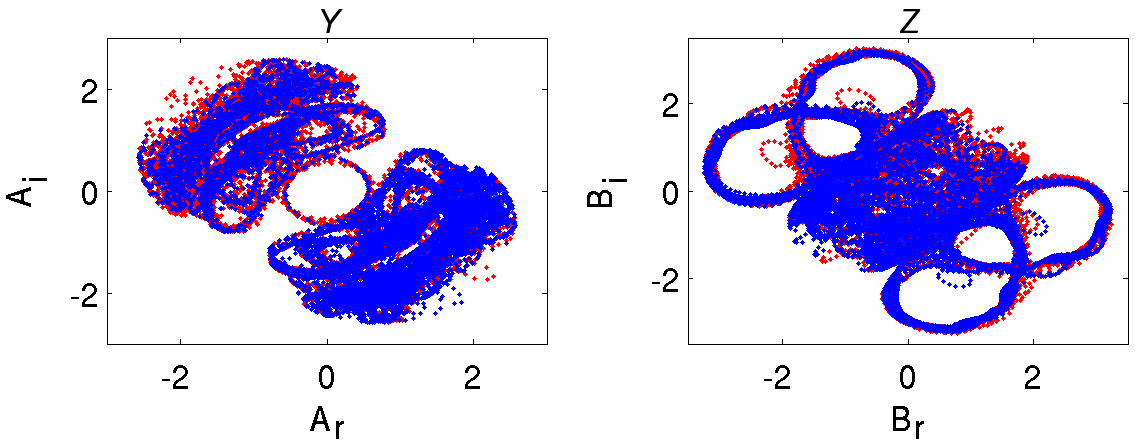}
\caption{(Color online) Chaotic whirling motion of the suspended carbon nanotube. Red and blue dots represent the dynamical flowing of the system for initial conditions in the first and second region of symmetry, respectively. In this case,  $\tilde{\gamma}=0.001$,  $\varepsilon=0.012$ , $\varkappa = 1.25$ ($L=$~\unit[1]{$\mu$m}, $h_z=$~\unit[40]{mT}, $R_{T}=$~\unit[53]{nm}).}
\label{nicechaos}
\end{figure*}

It turns out that for $\varkappa >  0.140$, the system leaves the regime of equilibria and enters into the one of periodic solutions. In fact, at the critical control parameter $\varkappa \sim 0.150$, there is an exchange of stability from the unique stable equilibrium to a stable limit cycle. This transformation in phase space is performed through a Poincar\'e-Andronov-Hopf bifurcation and the computed results show that the stable limit cycle grows in phase space for $0.150 \leq\varkappa \leq 1$. Hereupon, further changes in the control parameter will be reflected in the periodicity of the limit cycle as it can clearly be seen in Fig.~\ref{bifurcations}.  In this figure, the nanotube dynamics undergoes sucesive period-doubling cascade bifurcations in the amplitude modulation when varying $\varkappa$. Due to the symmetric character of the solutions, the results are only plotted in region $\mathtt{I}$.

Period-doubling bifurcations pave the way for chaotic dynamics \cite{argyris} and for the control parameter $\varkappa \sim 1.25$, a particular event takes place. In Fig.~\ref{nicechaos} the phase portraits for the nanowire amplitudes suggest that the initially disconnected symmetric regions $\mathtt{I}$ and $\mathtt{II}$ are now connected in a very complicated way. Furthermore, the flow diagrams for initial conditions in region $\mathtt{I}$ (red dots) and region $\mathtt{II}$ (blue dots) indicate that the system dynamically evolves in the two regions, \emph{i.e.}, the flow lines generated by Eqs.~\eqref{rwf} are sensitive to the initial values and are dense in both regions, where two strange attractors can be readily identified. Hence, the system is in the chaotic regime.

\section{Conclusions}
\label{final}

We have considered a voltage-biased nanoelectromechanical Josephson junction, where a suspended nanowire is serving as a weak link between two superconducting electrodes, in an inhomogeneous magnetic field. For our case study, we have assumed that the magnetic field is generated by an MFM cantilever tip and the nanowire is in the form of a metallic carbon nanotube. In such a scenario, the inhomegeneity of the field in conjunction with the Josephson current flowing through the tube, give rise to a Laplace force that induces the nanoresonator to perform a whirling movement. We have studied the time evolution of the amplitude and relative phase of this non-planar whirling motion and demonstrated that at the parametric resonance, their coupled dynamics exhibit a rich dynamical behavior characterized by: multistability, limit cycles and chaos. These stationary regimes depend on the degree of the magnetic field inhomegeneity, which in the present case, is related to the curvature radius of the magnetic cantilever tip.

The experimental implementation of the system considered in the article is plausible in light of current state-of-the-art nanofabrication techniques. For instance, J.-D. Pillet and collaborators designed and constructed a superconducting hybrid nanostructure that comprises a carbon nanotube suspended between two superconductors \cite{pillet}. Concerning the MFM cantilever tip, the experimental results reported by Victor N. Matveev and co-workers in Ref.~\cite{matveev} suggest that it is possible to fabricate cantilever tips coated by a magnetic film with a curvature radius in the range of $\unit[50-70]{nm}$ with  maximum magnetic fields in the range of $\unit[40-80]{mT}$. Due to a growing interest in complex behavior in nanodevices \cite{marcos}, the nonlinear and nonplanar phenomena exhibited by the system studied here have potential applications in signal processing, chaotic encryptation and random number generation \cite{conley, chen}.

\section{Acknowledgments}

The author wishes to thank Leonid Y. Gorelik for suggesting the problem, as well as to Gustav Sonne, Juan Atalaya, Tomasz Antosiewicz, Thomas Ericsson and  Mats Jonson for fruitful discussions and careful proof-reading. Financial support from the Swedish Research Council (VR) is gratefully acknowledged.

\bibliographystyle{apsrev}

\end{document}